\documentclass[aps,prl,reprint,superscriptaddress]{revtex4-2}

\usepackage{graphicx}
\usepackage{epstopdf}
\usepackage{amsmath,amssymb,amsfonts,bm}
\usepackage[inline]{enumitem}
\usepackage{verbatim}
\usepackage{mathdots}
\usepackage{xcolor}
\definecolor{apsblue}{RGB}{0,20,150}

\usepackage[
  colorlinks=true,
  linkcolor=apsblue,
  citecolor=apsblue,
  urlcolor=apsblue,
  pdfborder={0 0 0}
]{hyperref}

\pdfoutput=1

\begin{document}

\title{A Spectral Route to Directed-Polymer Glasses}

\author{Sen Mu}
\affiliation{Max Planck Institute for the Physics of Complex Systems, 01187 Dresden, Germany}

\author{Abbas Ali Saberi}
\email{asaberi@constructor.university}
\affiliation{School of Science, Constructor University, Campus Ring 1, 28759 Bremen, Germany}

\author{Mehran Kardar}
\affiliation{Department of Physics, Massachusetts Institute of Technology, Cambridge, Massachusetts 02139, USA}

\begin{abstract}
A finite density of mutually avoiding directed polymers in a quenched random medium is a minimal model of glassy line matter. The dilute theory, solved by replica Bethe ansatz, predicts an interaction free energy proportional to $\rho^2$ and disorder cumulants with distinct power-law dependences on the density $\rho$,
but direct numerical tests have been hindered by the combinatorially large many-polymer transfer matrix. We recast the problem as filling logarithmic eigenvalues of a single-polymer transfer-matrix product, obtaining the quenched free energy, its cumulants, and a disorder-induced linear spectral edge consistent with the replica prediction.
\end{abstract}

\maketitle

\textit{Introduction.---}
Directed polymers in random media (DPRM) provide a simple model of elastic matter configurations responding to a quenched disordered  landscape~\cite{KardarZhang87,HalpinHealyZhang1995,Zygouras2024}. 
They have shaped the theory of pinned domain walls~\cite{HuseHenley1985,Fisher1986} and are closely connected to stochastic interface growth~\cite{KPZ86}.
At finite density $\rho$, a collection of mutually avoiding DPRMs becomes a simple realization of glassy line matter: many paths optimize in the same random environment while being forbidden to cross. This setting is directly relevant to vortex lines in disordered type-II superconductors~\cite{Fisher1989,Blatter1994}, to non-crossing steps and discommensurations in two-dimensional systems~\cite{PokrovskyTalapov1979,Coppersmith1982,FisherFisher1982}, and to recent work on solvable and non-intersecting polymer ensembles~\cite{BorodinCorwinFerrari2014,OConnellWarren2016,DeLucaLeDoussal2015,DeLucaLeDoussal2016,BarraquandLeDoussal2023}.

In the dilute regime, the finite-density problem shows a sharp distinction between pure and disordered media. 
For mutually avoiding directed paths without disorder, at low density, the (entropic) interaction contribution to the free-energy density obeys the Pokrovsky-Talapov law $\Delta f(\rho)\sim \rho^3$~\cite{PokrovskyTalapov1979}.
By contrast, with quenched randomness, replica Bethe ansatz predicts (in the dilute continuum limit) a disorder-induced quadratic correction, $\Delta f(\rho)\sim \rho^2$~\cite{KardarNelson1985,Kardar1987},
together with corresponding low-density scaling forms for sample-to-sample free-energy cumulants~\cite{EmigKardar2000,EmigKardar2001}. 
Because these results rely on replica continuation and mappings to auxiliary many-body problems, direct numerical benchmarks are particularly desirable. However, while a single DPRM can be simulated efficiently by transfer-matrix methods~\cite{KardarZhang87,HuseHenley1985}, a direct many-polymer transfer matrix acts on a combinatorially large configuration space and becomes impractical at finite density.

Here we avoid the exponential growth of the many-DPRM transfer matrix by viewing mutually avoiding DPRMs as world lines of fermions evolving in imaginary time. For a fixed disorder realization, the single-polymer transfer-matrix product $W(t)$ is the one-particle propagator, while $m$ non-crossing polymers form an $m$-fermion state. At finite time the traced $m$-polymer partition function sums over all $m$-level fillings, and at long times is dominated by the product of the $m$ largest growth factors, so the free energy is governed by the leading logarithmic eigenvalues of $W(t)$. Thus the many-polymer glass can be studied by filling the spectrum of a single-polymer transfer-matrix product, rather than by constructing the combinatorially large many-polymer transfer matrix.
An early version of this approach was employed in~\cite{PolkovnikovKafriNelson2005} to explore pinning of vortex lines by columnar defects.

Mathematically, the free fermion construction is based on the determinant structure of non-intersecting paths, familiar from the Karlin--McGregor and Lindstr\"om--Gessel--Viennot formulas~\cite{KarlinMcGregor1959,Lindstrom1973,GesselViennot1985}. Its practical use, however, requires more than the usual Perron--Frobenius control of the leading eigenvalue. At finite density, the filling construction requires a stable set of positive real levels of the product matrix. This is achieved here by combining the total-nonnegative structure obtained from non-crossing path determinants, the oscillatory Jacobi structure of the transfer product, and high-precision stabilized multiplication.

Using this construction, we compute the quenched free-energy density from the filled logarithmic spectrum. The central numerical observation is that at low density, the mean cumulative growth per filled level decreases linearly with filling density, giving rise to the quadratic density dependence predicted by replica Bethe ansatz. Spectrally, this reflects a disorder-induced linear upper edge of the logarithmic band, in contrast to the quadratic edge of the usual (pure) free fermion problem.

\begin{figure}[t]
\centering
\includegraphics[width=0.48\textwidth]{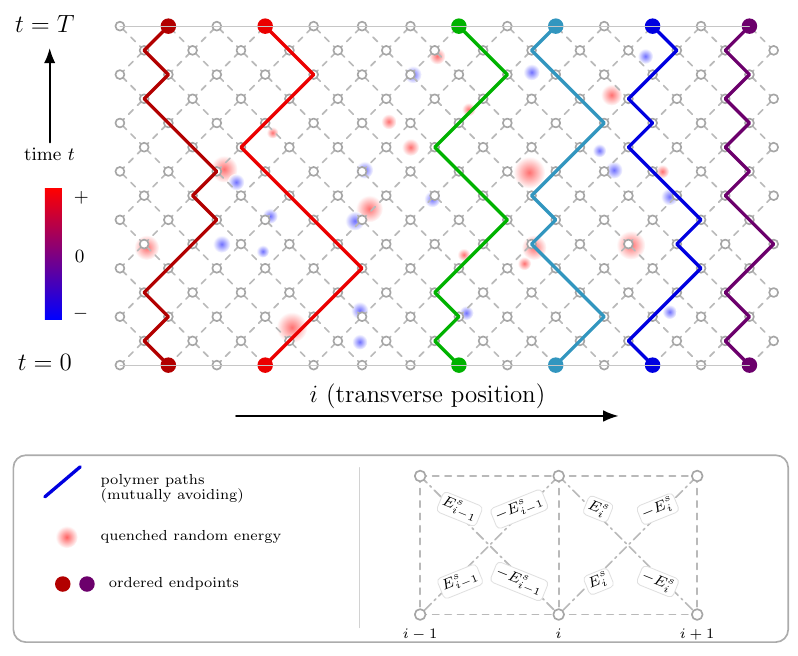}
\caption{
Finite-density mutually avoiding directed polymers in a quenched random medium.
Colored paths connect ordered endpoints at $t=0$ and $t=T$ on a space--time lattice with closed transverse boundary conditions, while the non-crossing constraint preserves their ordering.
Red and blue spots indicate the quenched disorder.
The lower-right block shows the bond-energy assignment defining $T(s)$, with the two sides of each local cell carrying $\pm E_i^{s}$ and the neighboring cell carrying $\pm E_{i-1}^{s}$.
}
\label{fig:directed_polymer_schematic}
\end{figure}

\textit{Model and spectral construction.---}
We use the lattice DPRM transfer-matrix ensemble introduced in Ref.~\cite{MuSaberiMoessnerKardar2026}. A polymer propagates along $\pm 45^\circ$ bonds of Fig.~\ref{fig:directed_polymer_schematic}, with the transfer matrix $T(s)$ constructed after two such steps. The 4 bonds around the intermediate vortex are chosen as $+E^s_i$ on one side and $-E^s_i$ on the other, with random bond energy $E^s_i$ drawn from a uniform distribution of mean $\mu$ and standard deviation $\sigma$.  With this choice the diagonal entries are $M^s_{ii}=\exp(-2E^s_{i-1})+\exp(2E^s_i)$, while all next-to-diagonal elements are set to 1.
This choice is somewhat arbitrary but ensures that the transfer matrix
reflects an underlying planar graph. The boundary elements are chosen consistently for the closed strip. The single-polymer evolution through one disorder realization is then governed by the product matrix
\begin{equation}
W(t)=T(t)T(t-1)\cdots T(1).
\label{eq:W_product}
\end{equation}
For a single polymer, endpoint choices correspond to matrix elements or contractions of $W(t)$. This observation was used in Ref.~\cite{MuSaberiMoessnerKardar2026} to realize distinct one-point fluctuation subclasses from different contractions of the same product matrix.

Here we use the product matrix for finite-density non-crossing DPRMs. For fixed ordered endpoints, the determinant construction gives the corresponding many-polymer weight as a minor of $W(t)$. The traced finite-density partition function is obtained by summing over ordered $m$-particle configurations. For one disorder realization, this free-fermion trace gives
\begin{equation}
Z_m(t)
=
\sum_{1\le i_1<\cdots<i_m\le N}
\lambda_{i_1}(t)\cdots\lambda_{i_m}(t),
\label{eq:Zm_fermion_sum}
\end{equation}
where the eigenvalues of $W(t)$ are ordered by $|\lambda_1|\geq|\lambda_2|\geq\cdots\geq|\lambda_N|$. In the long-time limit, the leading term is the product of the filled levels, giving the spectral filling rule
\begin{equation}
\ln Z_m(t)\simeq \sum_{i=1}^{m}\ln \lambda_i(t).
\label{eq:spectral_filling}
\end{equation}
Thus the traced many-polymer free energy is governed by the logarithmic spectrum of the single-polymer product.

The product $W(t)$ is generally non-symmetric, although each time-slice transfer matrix is real and symmetric. Perron--Frobenius theory therefore only guarantees that the leading eigenvalue is real and positive, whereas spectral filling requires real and positive eigenvalues throughout the filled sector. This property is not generic for entrywise positive matrices, but follows from the planar graph underlying the transfer matrix used here. The product $W(t)$ is the path matrix of the directed space--time network: its entries are single-polymer path sums through the time slab. For ordered initial and final sets $I=\{i_1<\cdots<i_k\}$ and $J=\{j_1<\cdots<j_k\}$, the Lindstr\"om--Gessel--Viennot/Karlin--McGregor theorem identifies the minor $\det W_{I,J}$ with the partition sum of $k$ mutually non-intersecting directed paths from $I$ to $J$ \cite{KarlinMcGregor1959,Lindstrom1973,GesselViennot1985}. The determinant signs implement the standard cancellation of intersecting path families, leaving only the ordered non-crossing families with positive sign. Since all microscopic Boltzmann weights are positive, these minors are nonnegative, giving the total-nonnegative structure needed for the filling construction.

Total nonnegativity alone, however, is not sufficient for an arbitrary non-symmetric matrix. The passage from nonnegative minors to real positive eigenvalues requires also the additional oscillatory structure of the nearest-neighbor transfer product, as detailed in Appendix A. In the Gantmacher--Krein framework, oscillatory matrices---nonsingular totally nonnegative matrices whose relevant ordered sectors become strictly positive under the dynamics---have real positive eigenvalues, with simplicity in the strictly oscillatory case~\cite{GantmacherKrein1950,Karlin1968,Ando1987}. This provides the spectral step needed for the filling rule, while the high-precision phase checks in Appendix B verify the real-positive filled sector over the time and density windows used in the data. Numerically, products are evaluated with stepwise rescaling, $W(t)=e^{C_t}A_t$, so that logarithmic eigenvalues are reconstructed as $\ln|\lambda_i(t)|=\ln|\tilde\lambda_i(t)|+C_t$, where $\tilde{\lambda}_i(t)$ denotes an eigenvalue of $A_t$ and $C_t$ is the accumulated overall rescaling factor; details of this stabilized reconstruction are given in Appendix B.

At density $\rho=m/N$, the quenched free-energy density is computed as
\begin{equation}
f_N(\rho,t)
=
-\frac{1}{Nt}\,\overline{\ln Z_m(t)}
\simeq
-\frac{1}{Nt}\,
\overline{\sum_{i=1}^{m}\ln\lambda_i(t)},
\label{eq:free_energy_density}
\end{equation}
where the overline denotes the quenched disorder average. The interaction part is obtained by subtracting the single-polymer contribution, $\Delta f(\rho)=f(\rho)-\rho f_1$, with $f_1=-\lim_{t\to\infty}t^{-1}\overline{\ln\lambda_1(t)}$.

\textit{Validation of the spectral filling method.---}
We next test the numerical stability of the filled logarithmic spectrum.
The corresponding precision check is shown in Figure~\ref{fig:validation_realness} of Appendix B. At working precision ${\tt wp}=3000$, where ${\tt wp}$ denotes the number of decimal digits retained in the high-precision arithmetic, apparent nonreal eigenvalues appear first only in the deepest part of the ordered spectrum. This is precisely where the rescaled logarithmic eigenvalues have become exponentially small and approach the numerical precision floor. Repeating the calculation at ${\tt wp}=16000$ resolves this deep tail over a much larger range and removes the apparent complex sector. The nonreal eigenvalues seen at lower precision are thus numerical artifacts of unresolved exponentially small modes, not intrinsic features of the filled logarithmic spectrum. We restrict the reported free-energy analysis to the stable filled sector.

A second check is available in closed form at full filling. When all $N$ levels are filled, the traced fermionic partition function is
\begin{equation}
Z_N(t)=\det W(t)=\prod_{i=1}^{N}\lambda_i(t).
\label{eq:full_det}
\end{equation}
Since determinants multiply, $\det W(t)=\prod_{s=1}^{t}\det T(s)$, for the present boundary convention, 
one has
\begin{equation}
\det T(s)=\exp\!\left[-\sum_{i=1}^{N}2E_i^s\right].
\label{eq:slice_det}
\end{equation}
This represents a fully packed zig-zag of paths with 
\begin{equation}
F_N(t)\equiv \ln Z_N(t)
=
-\sum_{s=1}^{t}\sum_{i=1}^{N}2E_i^s.
\label{eq:FN_exact}
\end{equation}
Since $E^s_i$ is drawn from a uniform distribution of mean $\mu=0$ and standard deviation $\sigma$, we have
\begin{equation}
\overline{F_N(t)}=0, \quad\frac{\overline{F_N^2(t)}_c}{Nt}=4\sigma^2 .
\label{eq:FN_variance_exact}
\end{equation}
This exact identity fixes both the mean of the filled logarithmic spectrum and the large-density fluctuation scale.

\begin{figure}[t]
    \centering
    \includegraphics[width=1.0\linewidth]{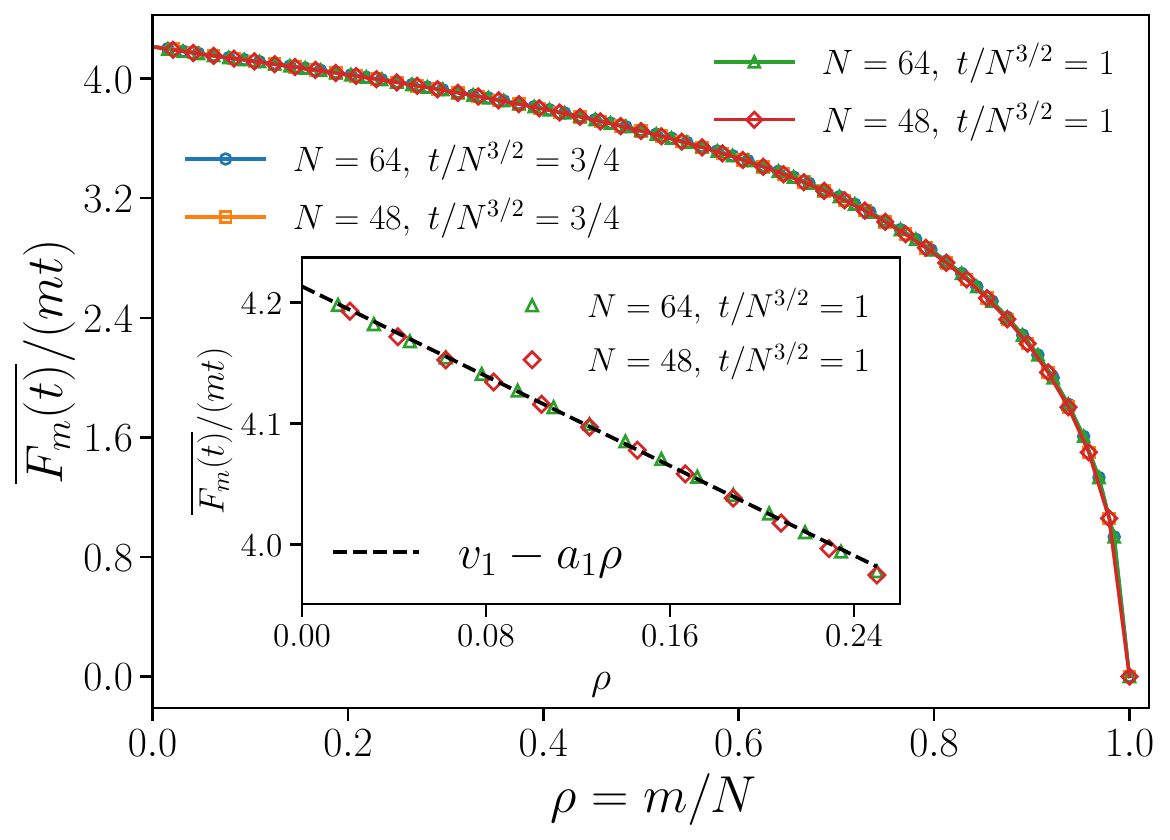}
    \caption{Mean cumulative growth per filled level, $\overline{F_m(t)}/(mt)$, as a function of the density $\rho=m/N$ for $N=48$ and $64$ at the matched scaled times $t/N^{3/2}=3/4$ and $1$. For $N=64$, these correspond to $t=384$ and $512$, while for $N=48$ the nearest integer times are $t=249$ and $333$. The close overlap of the curves at fixed $t/N^{3/2}$, as well as between the two scaled times, indicates an approximately stationary long-time regime. Inset: dilute-density regime $\rho\leq 1/4$ at $t/N^{3/2}=1$. The dashed line is a linear fit to the $N=64$ data, $\overline{F_m(t)}/(mt)=v_1-a_1\rho$, with $v_1\approx4.2$ and $a_1\approx0.9$; the $N=48$ data closely follow the same relation. The numerical simulations use $\mu=0$ and $\sigma=2$, with $14904$ disorder realizations for $N=64$ and $12384$ realizations for $N=48$.
    }
    \label{fig:free_energy_density_rho}
\end{figure}

\textit{Quenched free energy.---}
A natural finite-density descriptor is the cumulative logarithmic spectrum $F_m(t)=\sum_{i=1}^{m}\ln\lambda_i(t)$, since the $m$-polymer free energy fills the first $m$ logarithmic levels collectively. Individual levels or adjacent gaps probe local spectral structure, but do not by themselves define the finite-density free energy.

Figure~\ref{fig:free_energy_density_rho} shows  the disorder-averaged cumulative logarithmic growth per filled level, $\overline{F_m(t)}/(mt)$, as a function of density $\rho=m/N$. The curves for $N=48$ and $N=64$ nearly collapse at the matched scaled times $t/N^{3/2}=3/4$ and $t/N^{3/2}=1$, indicating that the normalized growth rate has reached an approximately stationary regime. In the dilute window, the inset shows that the data are well fitted to
\begin{equation}
\frac{\overline{F_m(t)}}{mt}
=
v_1-a_1\rho ,
\label{eq:linear_density_fit}
\end{equation}
where $v_1$ is the leading single-polymer logarithmic growth rate. 
Substituting Eq.~(\ref{eq:linear_density_fit}) into Eq.~(\ref{eq:free_energy_density}) gives
\begin{equation}
f_N(\rho,t)\simeq -\rho v_1+a_1\rho^2 ,
\label{eq:rho2_from_fit}
\end{equation}
so that, after subtracting the single-polymer term, $\Delta f(\rho)\sim \rho^2$. Thus the cumulative logarithmic spectrum directly exhibits the density dependence predicted by the replica Bethe ansatz. 

In the dilute continuum theory, the replica Bethe ansatz, together with a mapping to the dilute interacting Bose gas, predicts the density hierarchy $\overline{F_m^p(t)}_c/(Nt)= a_p \rho^{(5-p)/2}$, or equivalently $\overline{F_m^p(t)}_c/(mt)=a_p \rho^{(3-p)/2}$~\cite{EmigKardar2000,EmigKardar2001}. Here $\overline{(\cdots)}_c$ denotes a disorder cumulant. 
The above scalings have a simple physical interpretation: At density $\rho$, the typical distance between polymers is $\ell\sim 1/\rho$. Fluctuations of a single DPRM indicate a characteristic $\tau\sim \ell^{3/2}$ between collisions, and hence $\sim Nt/\rho^{-5/2}\sim (mt)\rho^{3/2}$ characteristic areas. For each area, fluctuations are expected to take the form $\rho^{-1/2}\chi$, where $\chi$ is likely a Tracy--Widom type random number with finite cumulants.
Related cumulant results for a single DPRM in finite transverse geometries are reported in Refs.~\cite{BrunetDerrida2000PRE,BrunetDerrida2000PhysicaA,BarraquandLeDoussal2025}.

\begin{figure}[t]
\centering
\includegraphics[width=1.0\linewidth]{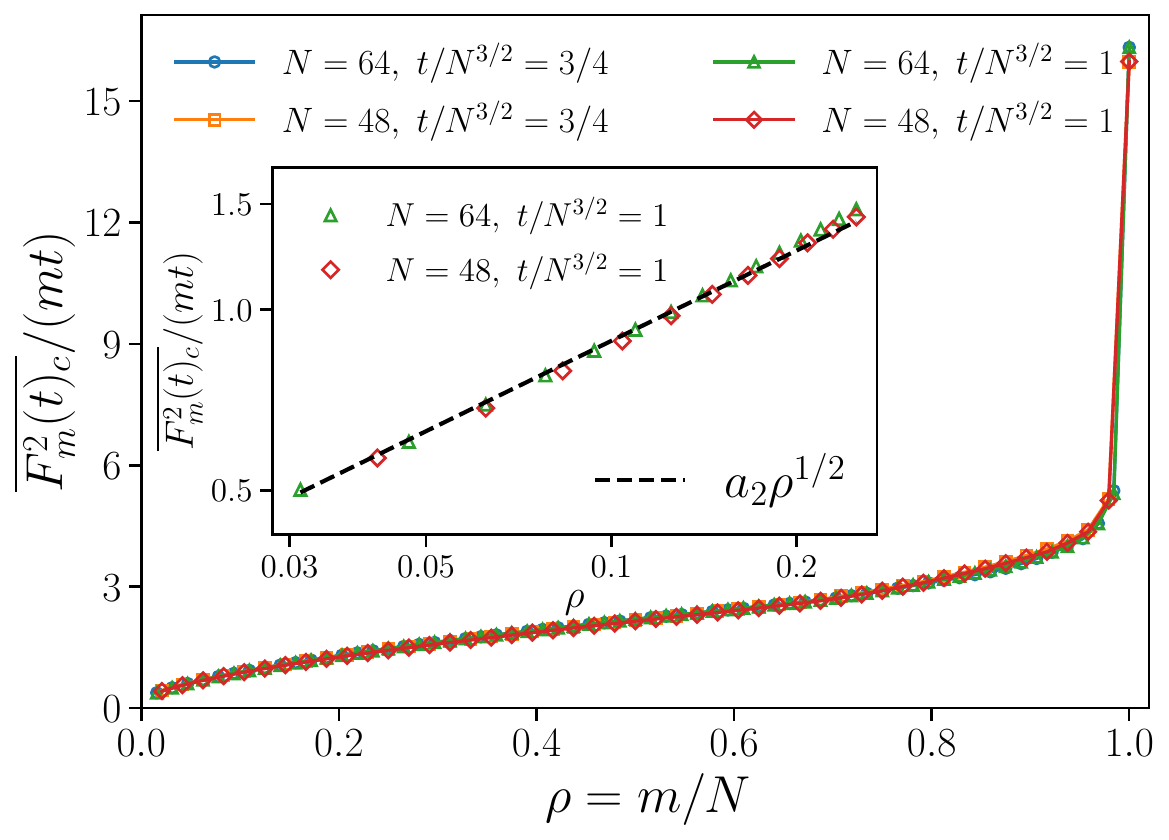}

\caption{Second disorder cumulant of the filled free energy, $\overline{F_m^2(t)_c}/(mt)$, as a function of the density $\rho=m/N$ for $N=48$ and $64$ at the matched scaled times $t/N^{3/2}=3/4$ and $1$. The sharp increase near full filling approaches the exact result $\overline{F_N^2(t)_c}/(Lt)=4\sigma^2=16$ for $\sigma=2$. Inset: log-log plot of the dilute regime at $t/N^{3/2}=1$. The dashed line shows the fit $a_2\rho^{1/2}$ to the $N=64$ data with $a_2\approx 2.9$; the $N=48$ data
closely follow the same scaling.
}
\label{fig:variance_free_energy_density_rho}
\end{figure}

Figure~\ref{fig:variance_free_energy_density_rho} shows the numerical results for the second cumulant. At small density, the inset is consistent with the scaling prediction $a_2\rho^{1/2}$. On approaching full filling, the data jump to the exact scale $\overline{F_N^2(t)}_c/(Nt)=4\sigma^2=16$ for $\sigma=2$, confirming that the rapid increase is anchored by the fully packed determinant identity rather than by numerical artifacts. 

Consistent with the scaling prediction, numerical data (not shown) indicate a third cumulant that starts at a finite value $a_3$ at low density, and decreases to 0 at $\rho=1$ (as dictated by the symmetry of the input random energies). 
The mapping to the dilute Bose gas~\cite{EmigKardar2001} goes beyond the scaling of the cumulants at low density, and provides exact values for the amplitudes, in terms of two model-dependent parameters. Indeed, from extracted values of $a_1$ and $a_2$ in Figs.~\ref{fig:free_energy_density_rho} and~\ref{fig:variance_free_energy_density_rho} there is a precise prediction for $a_3=\frac{9(12-\pi^2)}{128}(a_2^2/a_1)\approx 0.15(a_2^2/a_1)$. Given the statistical uncertainties, our current results are consistent with this prediction, but more statistics are required to provide a definitive test.

\textit{Spectral origin of the $\rho^2$ repulsion.---}
The quadratic density dependence has a direct spectral interpretation. Let
$\epsilon_i(t)=t^{-1}\ln\lambda_i(t)$ denote the ordered logarithmic levels for one disorder realization, with the one-based convention $i=1,\ldots,N$ used in the formulas. Filling all levels with $i\leq m$ gives
\begin{equation}
f_N(\rho,t)
\simeq
-\frac{1}{N}\,
\overline{
\sum_{i=1}^{\rho N}\epsilon_i(t)
}.
\label{eq:filled_spectrum}
\end{equation}
Thus, the interaction part of the free energy is controlled by how the upper edge of the disorder-averaged logarithmic spectrum deviates from its leading level.

\begin{figure}[t]
\centering
\includegraphics[width=1.0\linewidth]{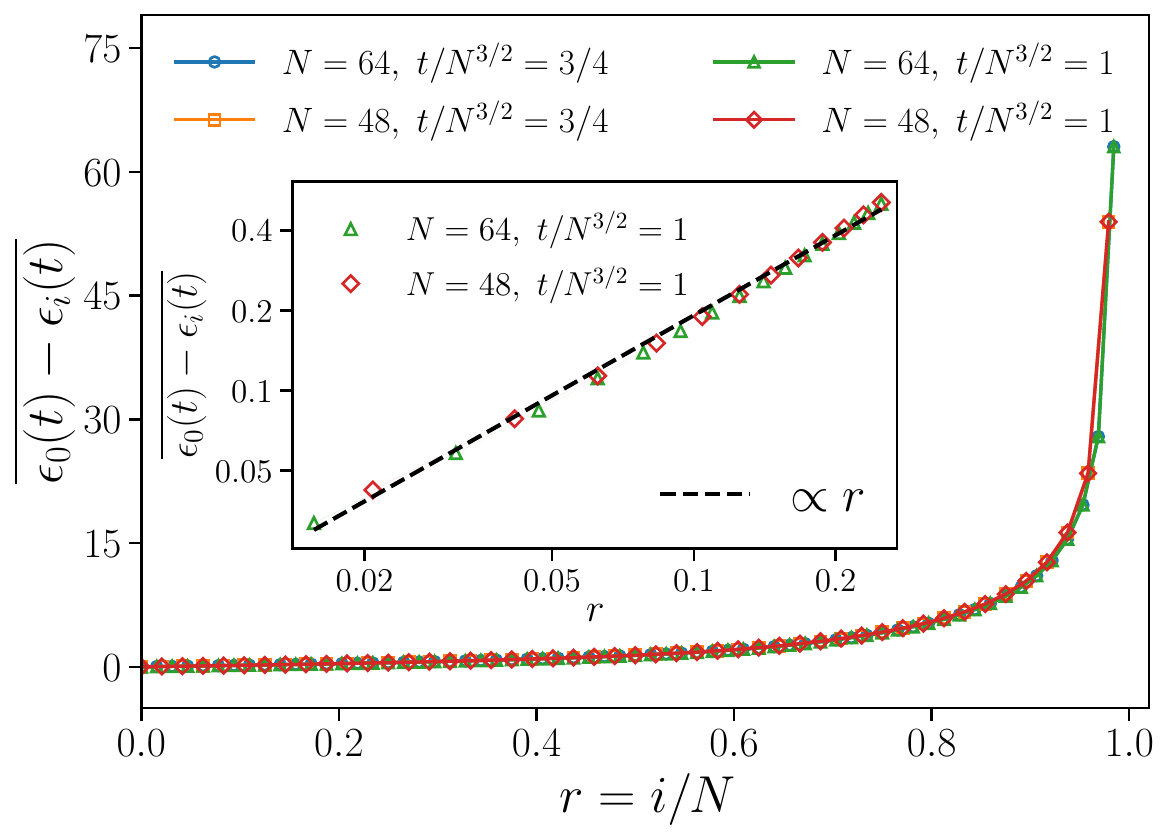}
\caption{
Rank-resolved logarithmic spectral edge. For this figure only, the levels are indexed as $i=0,\ldots,N-1$, with $i=0$ denoting the largest logarithmic eigenvalue, and $\epsilon_i(t)=\ln|\lambda_i(t)|/t$. The main panel shows
$\overline{\epsilon_0(t)-\epsilon_i(t)}$ as a function of the scaled rank $r=i/N$ for $N=48$ and $64$ at the matched scaled times $t/N^{3/2}=3/4$ and $1$. Inset: log-log plot of the small-$r$ regime $0<r\leq1/4$ at $t/N^{3/2}=1$. The dashed line is a linear fit to the $N=64$ data, demonstrating $\overline{\epsilon_0-\epsilon_i}\propto r$ near the upper spectral edge; the $N=48$ data closely follow the same scaling.}

\label{fig:spectral_edge}
\end{figure}

Figure~\ref{fig:spectral_edge} shows the rank-resolved edge directly. For this plot only, we use the zero-based rank convention $i=0,\ldots,N-1$, so that the largest logarithmic level is at $r=i/N=0$ without a finite-$1/N$ offset. If the continuum edge satisfies $\epsilon_0(t)-\epsilon_i(t)\propto r$ at late times, then filling the edge gives $\Delta f(\rho)\sim\int_0^\rho r\,dr\sim\rho^2$.
This edge behavior is distinct both from the corresponding pure free-fermion/Pokrovsky--Talapov case and from the conventional soft edge of real symmetric random-matrix theory. In the pure non-crossing problem, the one-particle band has a quadratic upper edge, $\epsilon_0-\epsilon_i\propto r^2$, giving $\Delta f_{\rm pure}(\rho)\sim\rho^3$. By contrast, for a standard Wigner--Dyson ensemble such as the GOE, the semicircle law gives a square-root density near the upper edge, $\rho_{\rm sc}(E)\sim(E_+-E)^{1/2}$ \cite{Wigner1958,Mehta2004}. If $r$ denotes the fraction of eigenvalues above energy $E$, then
$r\sim\int_E^{E_+}(E_+-E')^{1/2}\,dE'\sim(E_+-E)^{3/2}$, and hence the ordered edge obeys $E_+-E_r\sim r^{2/3}$. Filling such a conventional soft edge would therefore give an interaction scale $\int_0^\rho r^{2/3}dr\sim\rho^{5/3}$, with the largest-eigenvalue fluctuations governed by the Airy/Tracy--Widom edge statistics \cite{TracyWidom1994,TracyWidom1996}. Thus the observed edge is not a standard free-fermion or Wigner--Dyson edge, but a disorder-shaped logarithmic edge whose filling realizes the replica-Bethe-ansatz $\rho^2$ law.

\textit{Conclusion.---}
The main aim is to numerically study a finite-density glass of mutually repelling directed polymers. 
The non-crossing constraint turns the problem into a fermionic filling problem, without the need to construct a many-polymer transfer matrix.
The free energy is encoded in the ordered logarithmic spectrum of a single-polymer transfer-matrix product $W(t)$. 

The spectral perspective gives a simple interpretation of the replica-Bethe-ansatz results. In the pure system, the upper edge of the one-particle band is quadratic, and filling this edge gives the Pokrovsky–Talapov correction $\Delta f(\rho)\sim \rho^3$. In the disordered case, the upper edge of the logarithmic transfer-product spectrum is instead linear, resulting in $\Delta f(\rho)\sim \rho^2$ upon filling. The same filled spectrum also gives access to sample-to-sample fluctuations, whose low-density behavior is consistent with the predicted cumulant hierarchy, while the fully packed determinant identity anchors the opposite, unit-density limit.

Conversely, the physical system of non-crossing DPRMs leads to a mathematically interesting ensemble of random matrices $\{t^{-1}\ln W(t)\}$ whose properties are distinct from standard random matrices. 
The connection between non-crossing DPRM glass and spectral properties of this random matrix ensemble can be explored in several directions: testing the universality of the linear edge under changes of disorder and transfer-matrix ensemble, studying the crossover from dilute to fully packed densities, and exploring spectral signatures of vortex-line arrays, non-crossing disordered interfaces, and related line ensembles.

\textit{Acknowledgements.---}A preliminary version of this work appears in the PhD thesis of Sherry Chu at MIT~\cite{SherryChu}.
M.K. acknowledges discussions with Ramis Movassegh.
S.M. thanks J. Karcher
for helpful discussions on numerical simulations. This work was funded by the Deutsche Forschungsgemeinschaft (DFG, German Research Foundation) under Project No.~557852701 (A.A.S.). The study was also supported by the Advanced Study Group ``Strongly Correlated Extreme Fluctuations'' at the Max Planck Institute for the Physics of Complex Systems, Dresden (2024/25) \cite{pks_asg2024}, and the NSF through Grant No.~DMR-2218849 (M.K.).

\textit{Data availability.---}The data that support the findings of this article are not publicly available upon publication because it is not technically feasible and/or the cost of preparing, depositing, and hosting the data would be prohibitive within the terms of this research project. The data are available from the authors upon reasonable request.

\onecolumngrid
\vspace{1em}
\begin{center}
\textbf{End Matter}
\end{center}
\vspace{1em}
\twocolumngrid

The End Matter collects two technical ingredients used in the main text. Appendix A explains the positivity structure underlying the filled logarithmic spectrum, while Appendix B describes the stabilized high-precision reconstruction of the transfer-matrix product and presents the precision check for the filled sector.

\appendix

\noindent

\textit{Appendix A}: \textit{Spectral positivity of the transfer-matrix product---}
\label{app:positivity}
\setcounter{equation}{0}
\renewcommand{\theequation}{A\arabic{equation}}
We give the details behind the positivity statement used in the main text. The issue is that $W(t)$ is generally non-symmetric, so Perron--Frobenius controls only the leading eigenvalue, while spectral filling requires real positive eigenvalues throughout. The argument combines a path representation of minors with the oscillatory structure of the nearest-neighbor transfer product.

We first explain the minor positivity. For ordered row and column sets $I,J$ of the same size, a \emph{minor} is the determinant of the submatrix $A_{I,J}$. A matrix is \emph{totally nonnegative} when all such minors are nonnegative. This is the relevant notion here because a one-step nearest-neighbor transfer matrix is banded, so some minors vanish for geometrical reasons and strict positivity of all one-step minors is not expected. Total nonnegativity is stable under products. Indeed, the Cauchy--Binet formula gives
\begin{equation}
\det[(AB)_{I,J}]
=
\sum_{K:\, |K|=|I|}
\det(A_{I,K})\det(B_{K,J}) .
\label{eq:CB_appendix}
\end{equation}
Thus, if the one-step matrices have nonnegative minors, their product has the same property.

For the transfer matrix used in this work, these nonnegative minors have a direct physical origin. The product $W(t)=T(t)\cdots T(1)$ is the path matrix of the directed space--time network: the element $W_{ij}(t)$ is the partition sum of all single-polymer paths through the time slab from transverse point $j$ to transverse point $i$. Although the coarse-grained transfer-matrix description is written in terms of discrete time layers, each $T(s)$ represents the local vertex update across the elementary strip between two consecutive layers. In this elementary update, the diagonal matrix element is not an arbitrary onsite weight. Rather,
$M^s_{ii}=\exp(-2E^s_{i-1})+\exp(2E^s_i)$ is the local partition sum over two allowed positive vertex continuations between the two layers, while the off-diagonal entries represent allowed nearest-neighbor continuations with positive weight. This special Boltzmann construction, not entrywise positivity alone, is what makes the transfer matrix a positive path matrix.

The Lindstr\"om--Gessel--Viennot/Karlin--McGregor theorem then gives the path representation of these minors, turning each minor of the single-polymer path matrix into a multi-polymer partition function: $\det W_{I,J}$ equals the total weight of ordered, mutually non-intersecting directed path families from $I$ to $J$ \cite{KarlinMcGregor1959,Lindstrom1973,GesselViennot1985}. For ordered initial and final sets $I=\{i_1<\cdots<i_k\}$ and $J=\{j_1<\cdots<j_k\}$,
\begin{equation}
\det W_{I,J}
=
\sum_{\Gamma\in\mathcal{N}(I,J)}
\prod_{\gamma\in\Gamma} w(\gamma).
\label{eq:LGV_appendix}
\end{equation}
Here $\mathcal{N}(I,J)$ denotes the set of $k$ mutually non-intersecting directed path families from $I$ to $J$, and $w(\gamma)>0$ is the Boltzmann weight of one path. The determinant expansion itself contains signs; the content of the LGV/Karlin--McGregor theorem is that intersecting path families cancel pairwise by a sign-reversing construction. In the ordered planar geometry, the surviving non-intersecting families preserve endpoint order and therefore carry positive sign. Hence each relevant minor is a sum of positive weights. It is nonnegative in general and strictly positive whenever at least one compatible non-intersecting path family exists.

This positivity is special to the present vertex transfer matrix. A generic positive tridiagonal matrix need not have nonnegative minors: already a $2\times2$ block with positive diagonal entries $d_i,d_{i+1}$ and off-diagonal entries $1$ has determinant $d_i d_{i+1}-1$, which can be negative. By contrast, the Boltzmann form of our diagonal entries supplies precisely the positive local alternatives required by the path representation. This can be seen explicitly in the smallest nontrivial boundary block. Suppressing the time label and writing $q_i=\exp(2E_i)>0$, the $3\times3$ one-step block is
\begin{equation}
T_3=
\begin{pmatrix}
q_0^{-1}+q_1 & 1 & 0\\
1 & q_1^{-1}+q_2 & 1\\
0 & 1 & q_2^{-1}
\end{pmatrix}.
\label{eq:T3_appendix}
\end{equation}
Its leading determinants are $D_1=q_0^{-1}+q_1>0$ and
$D_2=(q_0^{-1}+q_1)(q_1^{-1}+q_2)-1
=q_0^{-1}q_1^{-1}+q_0^{-1}q_2+q_1q_2>0$; the full determinant is
$D_3=q_0^{-1}q_1^{-1}q_2^{-1}>0$.
The cancellation in $D_2$ and $D_3$ is the finite-dimensional trace of the vertex construction: the negative determinant terms produced by the off-diagonal couplings are compensated by positive alternatives contained in the diagonal Boltzmann sums. For a generic diagonal assignment, this compensation would not occur. Thus the sum-of-exponentials form is an essential local ingredient of the vertex-transfer construction, ensuring that the update between consecutive time layers realizes the non-crossing path positivity required for total nonnegativity.

It remains to connect total nonnegativity to the eigenvalues. Total nonnegativity alone is not a complete spectral theorem for an arbitrary non-symmetric matrix. The additional input is the oscillatory structure associated with the nearest-neighbor path matrix. In the terminology of Gantmacher--Krein, an oscillatory matrix is a nonsingular totally nonnegative matrix with enough connectivity that the relevant ordered sectors become strictly positive under the dynamics. Classical oscillatory-matrix theory then implies real positive eigenvalues, with simplicity in the strictly oscillatory case \cite{GantmacherKrein1950,Karlin1968,Ando1987}. In the present setting, the one-step transfer matrices have the Jacobi-type nearest-neighbor form with positive off-diagonal couplings, while the path representation gives the total-nonnegative structure of their products. This is the mathematical step that connects the nonnegative-minor structure, under the oscillatory assumptions above, to the real positive logarithmic levels required by the spectral filling rule. The high-precision phase checks in Fig.~\ref{fig:validation_realness} verify this property for the filled levels used in the free-energy analysis. 

The fully packed determinant identity used as the large-density check in the main text follows from the same structure by setting $m=N$: $Z_N(t)=\det W(t)$ and $\det W(t)=\prod_s\det T(s)$, leading to Eqs.~(\ref{eq:FN_exact}) and (\ref{eq:FN_variance_exact}).

\textit{Appendix B}: \textit{Stabilized spectral computation---}\label{app:numerics}
\setcounter{equation}{0}
\renewcommand{\theequation}{B\arabic{equation}}
\setcounter{figure}{0}
\renewcommand{\thefigure}{B\arabic{figure}}
For each disorder realization, we compute the transfer-matrix product recursively while removing its overall exponential scale. Starting from $A_0=\mathbb{I}$ and $C_0=0$, at each step $s\geq1$ we form $\widetilde W(s)=T(s)A_{s-1}$, define $\alpha_s=\|\widetilde W(s)\|_F$ where $\|\cdots\|_F$ denotes the Frobenius norm, set $A_s=\widetilde W(s)/\alpha_s$, and update $C_s=C_{s-1}+\ln\alpha_s$. It then follows recursively that
\begin{equation}
W(s)=e^{C_s}A_s,
\qquad
\ln|\lambda_i(s)|
=
\ln|\tilde\lambda_i(s)|+C_s .
\end{equation}
where $\lambda_i(s)$ and $\tilde{\lambda}_i(s)$ are the corresponding eigenvalues of $W(s)$ and $A_s$, respectively, and $C_s$ is the accumulated overall rescaling factor. This procedure separates the common exponential growth of the product from its relative spectral structure. Because the product $W(t)$ is generally non-normal, resolving small subleading eigenvalues is more delicate than in Hermitian problems. We therefore report only the matrix sizes and times for which the logarithmic spectrum are stable upon increasing the working precision. The high-precision calculations reported in the main text use this criterion to distinguish reliably resolved filled levels from the precision-limited deep spectral tail.

Figure~\ref{fig:validation_realness} shows a phase diagnostic for the eigenvalues of $A_s$. At working precision ${\tt wp}=3000$, nonreal eigenvalues appear only at late times and only in the deepest part of the ordered spectrum, $i\simeq N$. This is precisely the regime where the logarithmic eigenvalues become extremely negative and approach the numerical precision floor, as indicated by the inset. At the increased working precision ${\tt wp}=16000$, no nonreal sector is observed over the same time window. In short, only at sufficiently high working precision can we numerically confirm the real and positive character of the eigenvalue spectrum; complex eigenvalues observed at lower precision are instead numerical artifacts arising from unresolved exponentially small modes.

\begin{figure}[!b]
\centering
\includegraphics[width=\linewidth]{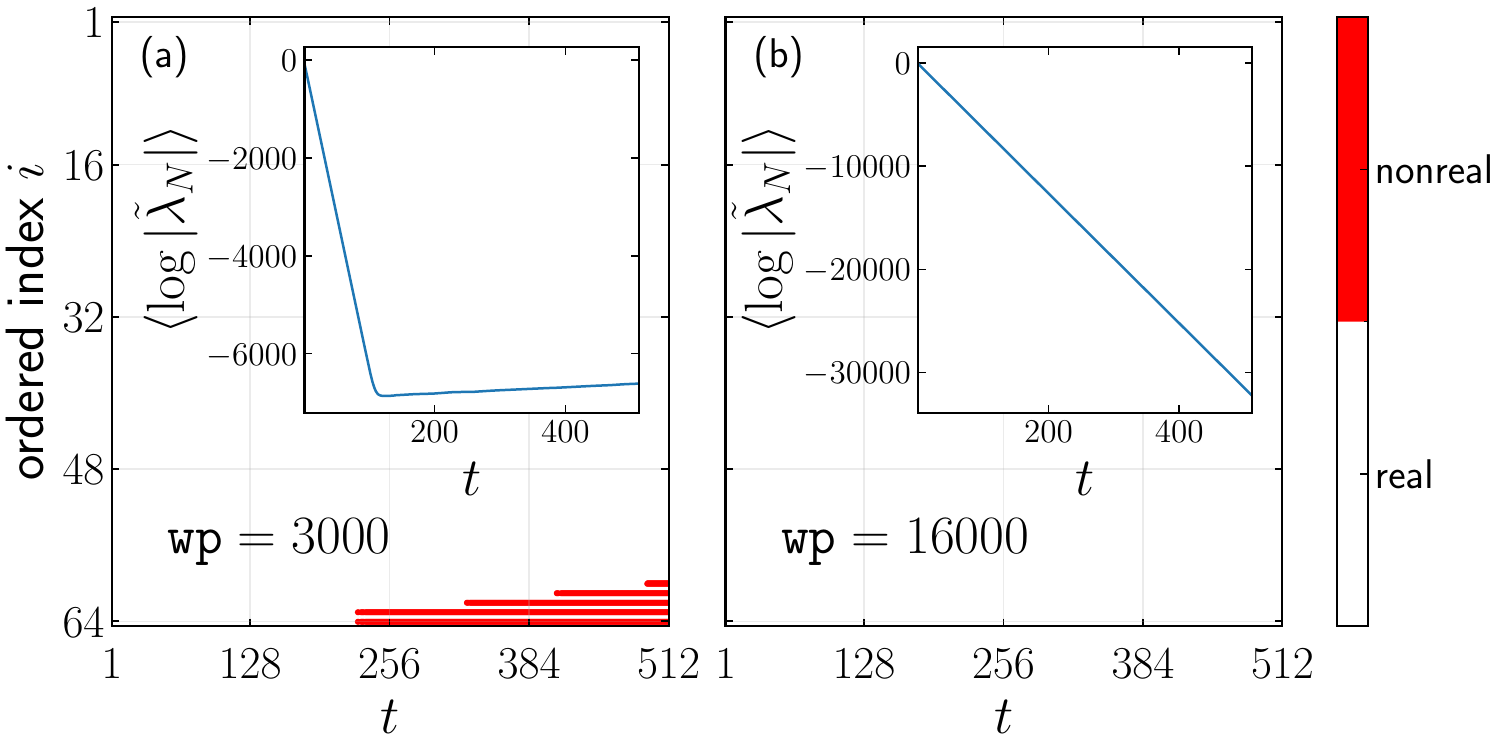}
\caption{
Precision check for the realness of the ordered spectrum of the rescaled transfer-matrix product $A_s$.
The two panels show the phase diagnostic for the rescaled eigenvalues at working precision ${\tt wp}=3000$ (a) and ${\tt wp}=16000$ (b) for $N=64$ up to $t=512$, where ${\tt wp}$ denotes the number of decimal digits retained in the high-precision arithmetic.
Red marks indicate $(t,i)$ points where at least one disorder realization contains a nonreal eigenvalue; white denotes real eigenvalues within the phase tolerance.
The disappearance of the red sector at ${\tt wp}=16000$ shows that the lower-precision complex sector is a precision-floor artifact of unresolved exponentially small modes. The insets show the disorder-averaged last rescaled logarithmic eigenvalue, $\langle \log |\tilde{\lambda}_N(t)|\rangle$, which tracks the smallest resolved mode in the ordered spectrum and therefore directly monitors proximity to the numerical precision floor.
}
\label{fig:validation_realness}
\end{figure}

\end{document}